\documentstyle[12pt]{article}
\begin{document}

\input epsf.tex


\newcommand{\nc}{\newcommand}
\nc{\beq}{\begin{equation}}
\nc{\eeq}{\end{equation}}
\nc{\beqa}{\begin{eqnarray}}
\nc{\eeqa}{\end{eqnarray}}
\nc{\lra}{\leftrightarrow}
\nc{\sss}{\scriptscriptstyle}
{\nc{\lsim}{\mbox{\raisebox{-.6ex}{~$\stackrel{<}{\sim}$~}}}
{\nc{\gsim}{\mbox{\raisebox{-.6ex}{~$\stackrel{>}{\sim}$~}}}
\def\dsl{\partial\!\!\!/}
\def\lameff{\lambda_{\rm eff}}
\def\Re{{\rm Re\,}}
\def\Im{{\rm Im\,}}
\def\ns{\!\!\!\!\!\!\!\!\!\!\!\!\!\!\!\!\!\!\!\!\!\!\!\!\!\!\!\!\!\!\!\!\!\!\!}
\def\NS{\ns\ns\ns\ns}
\def\dsl{\partial\!\!\!/}
\def\Pl{P_{\sss L}}\def\Pr{P_{\sss R}}
\def\VEV#1{\langle #1 \rangle}
\nc{\rd}{{\rm d}}
\def\bfD{{\bf D}}
\def\bff{{\bf f}}
\def\vn{\vec n}
\def\vs{\vec\sigma}
\def\vy{\vec y}
\def\sdy{\vec\sigma\cdot\vec y}
\def\sdya{\vec\sigma_a\cdot\vec y_a}
\def\vno{\vec n_0}
\def\vnp{\vec n_0'}
\def\bfI{{\bf 1}}
\def\vg{\vec g}
\def\vS{\vec\Sigma}
\def\Sdy{\vec\Sigma\cdot\vec y}
\def\bfP{{\bf P}}
\def\bfQ{{\bf Q}}
\def\tf{\tilde f}
\def\si{\bar s_i}
\def\sb{\bar s}


\begin{titlepage}
\pagestyle{empty}
\baselineskip=21pt
\rightline{McGill/95-33}
\rightline{CERN-TH-95/137}
\rightline{hep-ph/9506285}
\rightline{June, 1995}
\vskip .4in

\begin{center}
{\large{\bf Diffusion and Debye Screening\\
            Near Expanding Domain Walls}}
\end{center}
\vskip .1in

\begin{center}
James M.~Cline

{\it McGill University, Montreal, PQ H3A 2T8, Canada,}

and

Kimmo Kainulainen

{\it CERN, CH-1211, Geneve 23, Switzerland.}
\end{center}

\vskip 0.7in

\centerline{ {\bf Abstract} }
\baselineskip=20pt
\vskip 0.5truecm
We study the effect of Debye screening of hypercharge when a net fermion number
is reflected from a domain wall during a first order phase transition, which
may be relevant for electroweak baryogenesis.  We give a simple method for
computing the effect of screening within the diffusion approximation, whose
results are compatible with those of a more elaborate treatment based on the
Boltzmann equation.  Our formalism takes into account the differences in
mobility of different particle species.  We believe it is conceptually simpler
than other accounts of screening that have appeared in this context. Somewhat
surprisingly, we find that Debye screening can actually {\it enhance}
electroweak baryogenesis by a modest factor ($\sim 2$).
\end{titlepage}
\baselineskip=20pt

An important ingredient of the charge transport mechanism of electroweak
baryogenesis \cite{CKN1,JPT1} is the diffusion of net left-handed fermion
number into the false vacuum, after quarks and leptons reflect from expanding
domain walls during the electroweak phase transition.  This left-handed
asymmetry is what drives sphalerons to produce a baryon asymmetry. It is
desirable to understand how it propagates into the plasma because the more
efficiently it does so, the longer it can drive baryon production before
getting stopped by collisions and overtaken by the advancing wall. A number of
papers have examined the diffusion problem from various angles
{\cite{JC}-\cite{Far}}.

In the early days of the charge transport proposal, there were some worries
that the chiral asymmetry, because it carries hypercharge, might be prevented
from penetrating deep into the plasma due to Debye screening, since hypercharge
is an unbroken gauged charge \cite{Khleb}.  Indeed this would be the case if
there were only one species of fermion carrying both chirality and hypercharge.
 It was argued however that when there are many species, the hypercharge can be
screened without implying that chirality is also screened \cite{CKN2,JPT2}.  In
fact reference \cite{CKN2} concluded that Debye screening has no effect
whatever on the rate of baryon production (up to small effects due to the
difference between Fermi-Dirac and Bose-Einstein statistics.  In the present
work we reexamine the effects of screening and show that in fact it can have
some effect on the baryon violation rate, actually making it somewhat larger
than if one had neglected screening.  The reason for the discrepancy in
conclusions is that we allow for the various species to have different
diffusion coefficients, which leads to some particles being more efficient
screeners than others.   We show how to explicitly compute the effects of
screening by incoporating the gauge forces responsible for it into the
diffusion approximation.  In this formalism one can derive rather elegantly the
final values of particle asymmetries after screening and diffusion, starting
from some initial asymmetries produced at the domain wall.

Our starting point, similarly to references \cite{Khleb} and \cite{CKN2}, is
the Boltzmann equation, including force terms for long range interactions. For
each species of particle it is
\beq
        \left({\partial \over\partial t} + \vec v\cdot{\partial \over\partial
        \vec x} + y_i \vec E\cdot  {\partial \over\partial \vec x} \right) f_i
        = {\cal C}_i,
\label{Boltz}
\eeq
where $y_i$ is the hypercharge of the $i$th species, $\vec E$ is the field
strength and ${\cal C}_i$ is a collision term.  We take $f_i$ and $n_i$ below
to refer to the net density of a given species, that is particle density minus
antiparticle density. Integrating over momenta gives the usual continuity
equation,
\beq
        {\partial n_i\over\partial t} + \vec\nabla\cdot\vec J_i = 0.
\eeq
Although the screening force appears to have been lost, it reappears in the
diffusion equation when we make a derivative expansion of the current in terms
of the density, because one must also add an Ohm's law term to $\vec J$:
\beq
        \vec J_i = - D_i\vec\nabla n_i + \sigma_i\vec E,
\label{ohm}
\eeq
where $\sigma_i$ is the conductivity and $D_i$ the diffusion constant of the
$i$th species.

There is a simple relation between the conductivity and the diffusion
coefficient.  The latter can be expressed in terms of the mean interaction time
and the mean squared velocity (between collisions) as $D = \overline{v^2}
\tau/3$, which for a relativistic particle is simply $\tau/3$. The conductivity
on the other hand is the induced current divided by $|\vec E|$, and the induced
current is $n\bar v$, where $\bar v$ is the average velocity due to the biasing
caused by the electric field.  After a collision the electric field causes the
particle to change its momentum by $\bar p = y_i|\vec E|\tau$ in the time
before the next collision, and $\bar v = \bar p/p$.  The result is that
\beq
        \sigma_i = 3ny_iD_i/p = y_i D_i T^2\left\{\begin{array}{ll}
        0.33 & {\rm bosons}\\ 0.21 & {\rm fermions} \end{array} \right.
\label{sDrel}
\eeq

Now we want to imagine the situation where a macroscopically large domain wall,
moving with a velocity $v$ in the $z$ direction, is the source of the particle
densities.  Thus $\vec\nabla n$ becomes $\partial n/\partial z\equiv n'$. Since
we are interested in solutions of the form $n(z-vt)$, which would be static in
the rest frame of the wall, we can replace $\partial n/\partial t$ with $-vn'$.
Furthermore the divergence of the electric field is given by Gauss's law,
$\vec\nabla\cdot\vec E = \sum y_i n_i$.  So
\beq
        vn'_i + D_i n''_i - \sigma_i\sum_j y_j n_j = 0,
\eeq
or in vector notation on the space of species,
\beq
        v\vn' + \bfD \vn'' - (\vy\cdot\vn)\vs = 0,
\label{maineq}
\eeq
where $\bfD$ is the diagonal matrix with entries $D_i$.  This equation can be
readily solved using the Laplace transform,
\beq
        \vec N(s) = \int_0^\infty dz \,e^{-zs}\vn(z).
\eeq
Making the definitions
\beqa
        \vno &=& \vn(0); \phantom{tahavaha} \vnp = \vn'(0);\nonumber\\
        \bff &=& vs + \bfD s^2; \qquad \vS = \bff^{-1}\vs;\nonumber\\
        \vg &=& s^{-1}\vno + \bff^{-1}\bfD\vnp,
\eeqa
 equation
(\ref{maineq}) becomes  $\vec N - (\vy\cdot\vec N)\vS = \vg$.  But by dotting
$\vy$ into this equation one can solve for $\vy\cdot\vec N$ and therefore for
$\vec N$,
\beq
        \vec N = \left(\bfI + {\vS\otimes\vy\over 1-\Sdy}\right)\cdot\vg.
\label{soln}
\eeq
The solution for $\vn(z)$ can then be obtained from the inverse Laplace
transform
\beq
        \vn(z) = \oint{ds\over 2\pi i}e^{zs}\vec N(s),
\eeq
where the contour is taken to encircle the left half-plane (including an
infinitesimal vertical strip of the right half-plane).  To evaluate $\vn(z)$ we
need to identify the residues of the poles of (\ref{soln}) contained within the
contour.  Furthermore the initial values $\vno$ and $\vnp$ are related to each
other by the physical requirement that the trivial constant solution does not
exist and that the solution has the correct boundary condition $vn(0) = \vno$.
So one must eliminate $\vnp$ in favor of $\vno$ to express the the solution in
closed form.

Let us start with the simplest situation, in which there is only one kind of
particle.  Then
\beq
    N(s) = \left(1 + {\sigma y\over f - \sigma y}\right)\left({n_0\over s}
    + {Dn'_0\over f}\right),
\eeq which has poles at $s=0$, $s_0=-v/D$ and
$s_\pm = - v/2D \pm (v^2 + 4D\sigma y)^{1/2}/2D$. But the residues at $s=0$ and
$s=s_0$ vanish, and $s_-$ is the only one remaining within the contour.  The
resulting solution in $z$-space is $n(z) = n_0\exp(s_-z)$.  The effect of Debye
screening is evident in this solution because the penetration depth is smaller
than it would have been in the absence of the long-range force.  In fact we can
identify the Debye screening length as
\beq
        k_D^{-1} = (\sigma y/D)^{-1/2}.
\eeq
When we take the limit of no screening, $\sigma y = 0$, the solution of course
reduces to that which has been previously found, $n_0 \exp(-vz/D)$, showing
that the penetration length gets longer as the wall gets slower, as one would
expect from computing the time the wall takes to catch up with a diffusing
particle.

Next consider the case where there are many particle species, but they all have
the same diffusion coefficient, $D$. Again $\vec N(s)$ has poles at $s=0$,
$s=s_0=-v/D$ and $s=s_\pm$, but now $s_\pm$ is given by $s_\pm = -v/2D \pm (v^2
+ 4D\sdy )^{1/2}/2D$.   In this case none of the residues automatically vanish;
however the one at $s=0$ must be made to do so for physical consistency, since
it would give a spatially constant solution.  This gives the condition
\beq
        \bfP\cdot\left(\vno + {\bfD\over v}\vnp\right) = 0,
\label{cons}
\eeq
where $\bfP$ ($\bfQ$) is a projection operator onto the subspace orthogonal
(parallel) to hypercharge,
\beq
        \bfP \equiv \bfI - \bfQ = \left(\bfI - {\vs\otimes\vy\over\sdy}\right).
\label{proj}
\eeq
Furthermore we get a relation between $\vno$ and $\vnp$ from the requirement
that $\vn(z)$ indeed gives $\vno$ at $z=0$,
\beq
        \bfQ\cdot\vnp = s_-\bfQ\cdot\vno.
\eeq
Thus the complete solution can be expressed in terms of $\vno$,
\beq
        \vn(z) = \left(e^{s_0z}\bfP + e^{s_-z}\bfQ\right)\cdot\vno,
\label{simple}
\eeq
which has a quite simple geometric interpretation: there is only screening in
the direction of particle species space proportional to hypercharge.

In the foregoing examples one could arrive at the solutions without recourse to
the machinery of Laplace transforming, but in the general case where there are
several diffusion coefficients, the solutions would be harder to guess.  We
assume that there are $M$ groups of particles classified by the values of their
diffusion coefficients $D_a$ (for example different flavors of quarks have
approximately  equal diffusion constants because of their interactions with
gluons).  It is useful to define operators analogous to (\ref{proj}) that are
nonzero only in the subspace of species with a given value $D_a$:
\beq
        \bfP_a \equiv \bfI_a - \bfQ_a = \left(\bfI_a - {\vs_a\otimes\vy_a
        \over\sdya}\right),
\eeq
where
\beqa
        \bfI_a &\equiv& \lim_{\epsilon\to 0}\ i\epsilon\, (\bfD - D_a +
        i\epsilon)^{-1};        \nonumber\\
        \vs_a &=& \bfI_a\cdot\vs;\qquad \vy_a = \bfI_a\cdot\vy.
\eeqa
There are $2M$ physical solutions for the poles of $\vec N(s)$ which lie in the
left-half plane.  $M$ of these are obvious generalizations of the previous
unscreened solution,
\beq
        s_a = -v/D_a.
\eeq
The other $M$ physical poles, which we shall call $\si$, occur at the negative
roots of an $(M+1)$th order equation,
\beq
        \sb\prod_a(\sb D_a+v) -\sum_a\sdya\prod_{b\neq a}(\sb D_b+v) = 0
\label{extras}
\eeq
The complete solution can be written as
\beq
        \vn(z) = \sum_{a=1}^Me^{s_az}\bfP_a\cdot\vno + \sum_{a,i,b=1}^M
        \vs_a {M_{ai}e^{\si z}(M^{-1})_{ib}\over
        \vs_b\cdot\vy_b}\vy_b\cdot\vno
\label{result}
\eeq
where $M_{ai}$ is a square matrix with elements
\beq
        M_{ai} = {1\over \si^2 D_a +\si v}
\eeq
In practice it is not $\vno$ but rather the fluxes $\vec J(0)$ at the boundary
which are known at the outset.  However it is straightforward to prove the
intuitive result that
\beq
        \vec J(0) = v\vno
\eeq
from eqs.~(\ref{ohm}) and (\ref{result}) using the identities $D_a M_{ai}\si =
\si^{-1} - vM_{ai}$, eq.~(\ref{extras}), and the electric field $|\vec E(0)| =
\int_0^\infty \vy\cdot\vn\, dz$.

It is trivial to verify that (\ref{result}) satisfies the original differential
equation; the nontrivial part is the determination of the roots $\si$.  If we
assume that $v^2/D\ll\sdy$, which is the condition that the Debye screening
length is much shorter than the usual diffusion length (and is always satisfied
by the domain walls of the electroweak phase transition), it can be shown that
two of the solutions to (\ref{extras}) are given by
\beq
        \pm (\sdya/D_a)^{1/2} + O(v/D),
\label{large}
\eeq
of which the negative one is physical and gives rise to a single direction in
the space of species which is screened.  In the same regime of small
velocities, the remaining $M-1$ solutions are approximately the roots of the
truncated version of eq.~(\ref{extras}) in which the first term (being of
higher order in $v$) is omitted.  This gives $M-1$ values of $\si$ which are
all of order $v/D$ and therefore correspond to unscreened contributions.

To illustrate the full solution we will consider a few special cases of
interest.  First let us assume that there are just $M=2$ different diffusion
coefficients, which could be useful for comparing the transport of quarks
versus leptons, for example.  Using eq.~(\ref{sDrel}) and ignoring the
difference between the bosons and fermions, we find the simple result
\beq
        \vn(z) = \left( \sum_{a=1}^2 e^{-vz/D_a} \bfP_a
        + {e^{\bar s z}\over\vy\cdot\vy}\left(
        R_2\vy_1 - R_1\vy_2\right)\otimes \left(\vy_1/R_1
        - \vy_2/R_2\right)\right)\cdot\vno,
\label{M2case}
\eeq
where
\beq
        \bar s = -{v\sdy\over D_1 \vs_2\cdot\vy_2 + D_2 \vs_1\cdot\vy_1}
         = -{v\over \vy\cdot\vy}\left({\vy_1\cdot\vy_1\over D_2} +
        {\vy_2\cdot\vy_2\over D_1}\right),
\label{twocpt}
\eeq
and
\beq
        R_a = {\vy_a\cdot\vy_a\over
         \vy\cdot\vy}.
\label{Rs}
\eeq

We have omitted the screened contribution here, which is concentrated much
closer to the bubble wall than the unscreened parts. It can be checked that
this result reduces to our previous one (\ref{simple}) in the limit that
$D_1\to D_2$.  It is not correct in the limit that one of the $D_a$'s vanish,
which corresponds to the decoupling of one of the particles, in the sense that
it becomes immobile.  This is because our assumption that $v^2/D\ll\sdy$ is
then violated, but since we expect this condition to be satisfied in nature
even for quarks, which have the smallest diffusion length, eq.~(\ref{twocpt})
is still the relevant one for considering the limit when one of the $D_a$'s is
much smaller than the other one.

The application motivating us to find this result was the case of differential
reflection between chiral fermions and antifermions from the advancing domain
wall as a means for producing the baryon asymmetry during the electroweak phase
transition.  In this mechanism, the net production of baryon number is
proportional to the integrated density of $n_{B_L}+n_{L_L}$, a linear
combination of the left-handed baryon and lepton number densities.  We can
write this as
\beq
        \Delta B = c \int_0^\infty dz\ \vec u\cdot\vn(z),
\label{bviol}
\eeq
where $\vec u$ is a vector with components $u_i = 1$ for every left-handed
particle and zero for all others, and $c$ is some constant. It is easy to
verify that $\vec u\cdot\vy = 0$, which stems from the fact that the baryon
violation interaction giving rise to (\ref{bviol}) is invariant under
hypercharge gauge transformations.  In the case of all equal diffusion
coefficients $D_i$, this has the consequence that $\Delta B \propto \vec u\cdot
\vno$ because $\vec u$ is orthogonal to hypercharge,
\beq
        \vec u\cdot\bfP\cdot\vno = \vec u\cdot\vno.
\eeq
This is the origin of the statement that hypercharge screening has no effect on
the baryon asymmetry:  we would have gotten the same answer if we had ignored
screening altogether.

However when the differences between diffusivity of particle species are taken
into account, it is no longer true that hypercharge screening has no effect
because, for example,
\beq
        \vec u \cdot\bfP_a\cdot\vno \neq \vec u \cdot \bfI_a\cdot \vno,
\eeq
because $\vec u$ is orthogonal to the full vector of hypercharges $\vy$, but
generally not to its projections $\vy_a$ onto the space of species with
diffusion coefficient $D_a$.

Let us next examine the case where there are $M=3$ distinct diffusion
coefficients with the hierarchy $D_1 \ll D_2 < D_3$, which is the case for
quarks and left-handed and right-handed leptons, respectively.  Since $D_1$ is
small we can expand in it.  To zeroth order in $D_1$, the integrated densities
in front of the wall are given by
\beq    \int_0^\infty dz\, \vn(z) =
        \left(\sum_{a=2}^3 D_a\bfP_a +
        {(R_3\vy_2 - R_2\vy_3)\otimes(\vy_2/R_2-\vy_3/R_3)
        \over \vy_2\cdot\vy_2/D_3 + \vy_3\cdot\vy_3/D_2}\right)
        \cdot\vno,
\eeq
which is just the same result as (\ref{M2case}): we recover the results for
$M=2$ when one of the particles becomes immobile ($D_1\to 0$). The $O(D_1)$
corrections which involve the $\vy_1$ direction either in the initial or the
final density are
\beq
        D_1\left( \bfP_1 +
        {(\begin{array}{ccc} \vy_1 &\vy_2 &
        \vy_3 \end{array})\over\bar D\vy\cdot\vy}
        \left(\begin{array}{ccc}
        \bar D & \Delta D R_2 R_3 & -\Delta D R_2 R_3 \\
        D_2(R_1-1)/R_1 & * & * \\
        D_3(R_1-1)/R_1 & * & * \end{array}\right)\otimes
        \left(\begin{array}{c}{\vy_1/R_1} \\{\vy_2/R_2} \\
        {\vy_3/R_3}\end{array}\right)\right)\cdot\vno,
\label{approx}
\eeq
with
\beq
        \bar D = R_2 D_2 + R_3 D_3;\qquad \Delta D = D_3 - D_2,
\eeq
and the $R_a$'s defined as in eq.~(\ref{Rs}). {}From (\ref{approx}) we see that
any particle densities induced by screening from an initial density of quarks
(proportional to $\vy_1\cdot\vno$ if we identify quarks with the 1-direction),
as well as any quark density induced from initial densities of leptons
($\vy_{2,3}\cdot\vno$) by screening, propagates only with the diffusion
coefficient $D_1$ of the quarks, which is much smaller than that of the
leptons.  Although this result may be intuitively obvious, the exact
proportions by which these contributions are suppressed were not.

To obtain more concrete results, we need to specify the values of the
hypercharges and the diffusion coefficients. The space of species consists of
18 left-handed quarks, 9 right-handed up-type quarks, 9 right-handed down-type
quarks, 6 left-handed leptons, 3 right-handed leptons, and let us say $2$ Higgs
doublets, counting colors and flavors.  We will take the Higgs bosons to have
the same diffusion coefficient as the left-handed leptons. The hypercharge
vector can thus be written as
\beqa
        \vy &=& (\vy_q;\ \vy_l;\ \vy_r)    \nonumber\\
        \vy_q &=& \left(18\times{1\over 3};\ 9\times{4\over 3};\
         9\times-{2\over3}\right);
             \qquad \vy_q\cdot\vy_q = 22;\nonumber\\
        \vy_l &=& \left(6\times{-1};\ 2\times{-1}\right);
        \phantom{hannatyttoi}\;\;
                    \vy_l\cdot\vy_l = 8;\nonumber\\
               \vy_r &=& \left(3\times{-2}\right);
        \phantom{IsinOmahannatytto}\;\;
               \vy_r\cdot\vy_r = 12.
\eeqa
Consider again the example with $M=2$, where we assume that right and left
handed leptons have equal diffusion coefficients (so that $\vy_{\rm
lepton}=\vy_l \oplus \vy_r$).  Then assuming that the inital flux consisted
only of left handed bottom quarks with total density $n_q$ and right handed
bottom quarks with density $-n_q$ and similarly for leptons with density $\pm
n_l$, we would find
\beq
\Delta B  =  {cv^{-1}}(n_qD_q + n_lD_l)
           - 3cv^{-1}{D_l - D_q\over 11D_q + 10D_l}
          \left( D_qn_q - D_ln_l \right).
\eeq
This result implies an enhancement in the flux due to the hypercharge screening
by about $30$ per cent.  It also illustrates the fact that regardless of
screening, the particles with the largest diffusion coefficient are still the
most efficacious for producing baryon number.

The most interesting physical application is the situation where the initial
state at the wall consisted only of left-handed tau leptons with density
$n_\tau$ and an equal and opposite density $-n_\tau$ of right-handed tau
leptons (since the process of reflection of fermions from the wall conserves
lepton number)~\cite{JPT1,CKV}. Neglecting the quark contribution entirely, we
then find that
\beq
   \Delta B = {cv^{-1}}{D_l \over 4}
              \left( 1  + {7\over 1+(2D_l)/(3D_r)} \right)
             \simeq {7\over 4}{cv^{-1}}D_l,
\label{analytic}
\eeq
using that the ratio of diffusion constants of left- and right-handed leptons
is approximately $D_l/D_r\cong 1/4$. The naive answer ignoring screening is
$\Delta B = cD_l/v$.  We thus see again the curious effect that screening {\it
increases} the net amount of propagation of left-handed lepton number into the
plasma.

\epsfysize7.5truecm\epsfbox{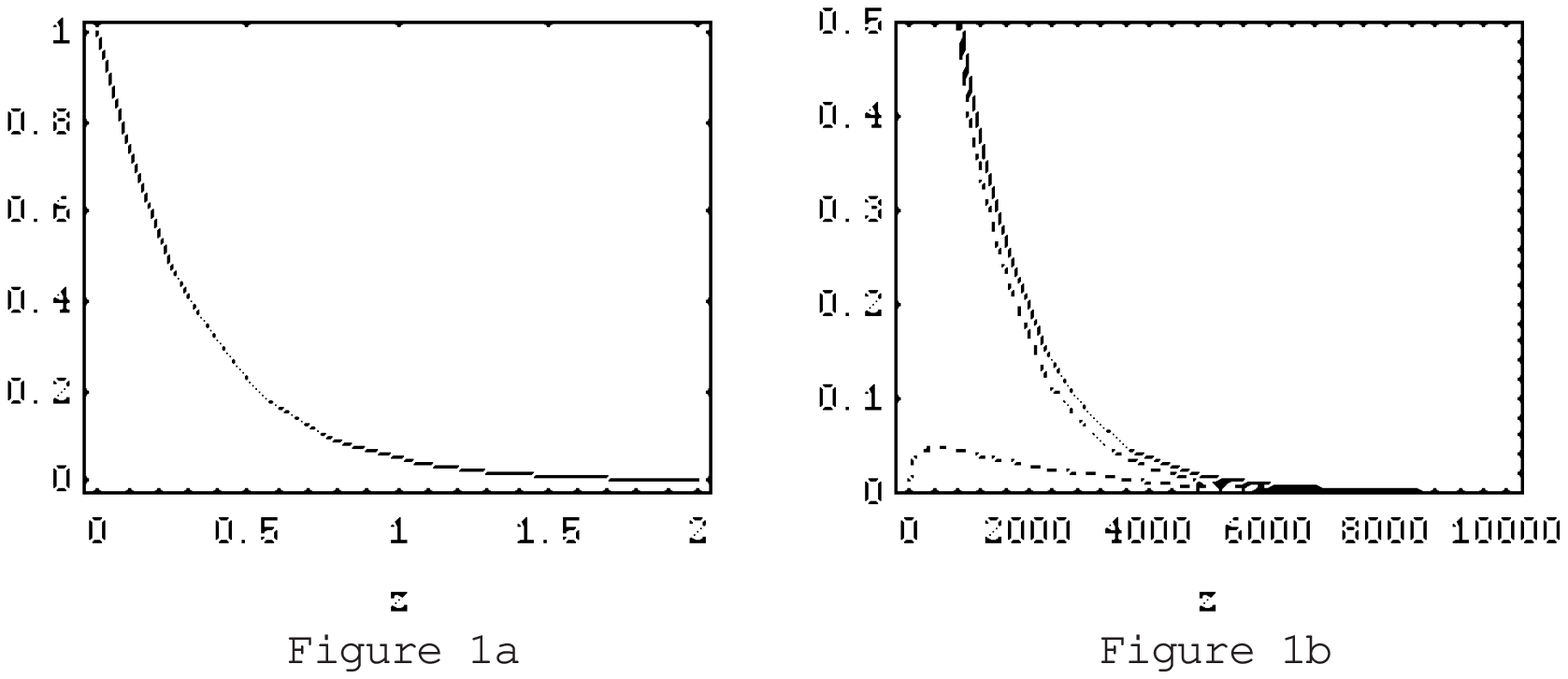}
\baselineskip 17pt
\noindent{\bf Figure 1.} Figure 1a shows the hypercharge density in front of
the wall. Figure 1b shows the spatial distribution of the asymmetries of the
left handed fermions. The solid line corresponds to the numerical solution of
equation (\ref{result}) for the $\tau_L$-asymmetry in the presence of screening
and the short dash - long dash one to the $\tau_L$-asymmetry in the absence of
screening. The dashed line  represents the $e_L,\mu_L$ and
$\nu_{iL}$-asymmetries ($i=e, \mu, \tau$), which would be zero in the absence
of the hypercharge screening.  Distance is in units of $T^{-1}$.
\vskip 0.5truecm

\baselineskip 20pt

We show some results from the full numerical solution of the equation
(\ref{result}) corresponding to the above case of initial tau fluxes only, and
with $D_q = 6/T$, $D_l = 110/T$ and $D_r = 4D_l$.  Figure 1a shows the
hypercharge density in front of the wall. The width of the distribution is seen
to correctly correspond to the inverse of the eigenvalue (\ref{large}). Figure
1b shows the distributions of the asymmetries of left handed leptons. Note in
particular the distributions of the  $e_L,\mu_L$ and neutrino-asymmetries that
the hypercharge screening pulls out from the plasma. It is the appearance of
these new leptonic asymmetries, along with the fact that the $\tau_L$ asymmetry
is only slightly reduced by the screening, that explains the enhancement in the
baryon production found above in the analytic formula (\ref{analytic}). The
precise numerical result for the final baryonic asymmetry is close to the
analytic estimate: $\Delta B \simeq 1.69 {cv^{-1}}D_l$.

Let us finally caution the reader that there may be problems with the
fundamental assumptions in the diffusion approach; Fick's law implicitly
assumes that the reflected asymmetry current has a near-thermal distribution,
whereas the initial flux emanating from the wall is rather strongly
concentrated at low momenta \cite{CKN1,JC,CKV}. It remains to be seen how much
this will affect the above results of the diffusion computation.

In conclusion, we have studied the diffusion of a reflected fermion current
from the domain wall during a first order phase transition, which is relevant
for  the charge transport mechanism of electroweak baryogenesis.  In our
treatment we included the effect of Debye screening of hypercharge and
accounted for the differences in mobility of different particle species.  Our
main result is that in the case where the baryogenesis is driven by the flux of
reflected tau leptons, Debye screening actually {\it enhances} electroweak
baryogenesis by a modest factor $\sim 1.7$.


\end{document}